 \documentclass[fleqn,twoside]{article}
 \usepackage{gc,amssymb,cite}

\def\barr{\left(\begin{array}}
\def\earr{\end{array}\right)}
\renewcommand{\beq}[1]{\begin{equation}\label{#1}}
\renewcommand{\bear}[1]{\begin{eqnarray}\label{#1}}

\eqsection

\newcommand{\R}{ {\mathbb R} }

\newcommand{\p}{\partial}

\begin{document}
\twocolumn[ \prepno{gr-qc/yymmnnn}{\GC{13} 137 (2007)}


\Title {An $S$-brane solution with acceleration\yy
     and small enough variation of $G$}

\Aunames{J.-M. Alimi\auth{1,a},
    V.D. Ivashchuk\auth{2,b,c} and V.N. Melnikov\auth{3,b,c}}

\Addresses{
\addr a {Laboratoire de l'Univers et de ses Theories
    CNRS UMR8102, Observatoire de Paris  92195, Meudon Cedex, France }
 \addr b {Centre for Gravitation and Fundamental Metrology,
        VNIIMS, 46 Ozyornaya St., Moscow 119361, Russia}
\addr c {Institute of Gravitation and Cosmology,
        Peoples' Friendship University of Russia,
    6 Miklukho-Maklaya St., Moscow 117198, Russia} }


\Abstract
{An $S$-brane solution with two non-composite electric branes and a set of
$l$ scalar fields is considered. The intersection rule for branes
corresponds to the Lie algebra $A_2$. The solution contains five factor
spaces with the fifth one  interpreted as ``our'' 3-dimensional space.  It
is shown that there exists a time interval where accelerating expansion of
``our'' 3-dimensional space is compatible with small enough value of
effective gravitational ``constant'' variation.}


]
\email 1 {jean-michel.alimi@obspm.fr}
\email 2 {ivas@vniims.ru}
\email 3 {melnikov@phys.msu.ru}

\section{Introduction}

As is well known \cite{Mel2,Mel,Mel3,3,4,Solv,IM1}, cosmological models in
scalar-tensor and multidimensional theories are a framework for describing
possible time variations of fundamental physical constants due to scalar
fields which are present explicitly in STT or are generated by extra
dimensions in multidimensional theories. In \cite{2}, we have obtained
solutions for a system of conformal scalar and gravitational fields in 4D
and calculated the presently possible relative variation of $G$ at the level
of less than $10^{-12} yr^{-1}$.

Later, in the framework of a multidimensional model with a perfect fluid and
two factor spaces (our 3D space of Friedmann open, closed and flat models
and an internal 6D Ricci-flat space) we have obtained the same limit for
such variation of $G$ \cite{BIM1}.

We have also estimated the possible variations of the gravitational
constant $G$ in the framework of a generalized (Bergmann-Wagoner-Nordtvedt)
scalar-tensor theory of gravity on the basis of field equations, without
using their special solutions. Specific estimates were essentially related
to values of other cosmological parameters (the Hubble and acceleration
parameters, the dark matter density etc.), but the values of $\dot G/G$
compatible with modern observations did not exceeded $10^{-12}$ per year
\cite{BMN-G}.

In \cite{MIWaseda}, we continued the studies of models in arbitrary
dimensions and obtained relations for $\dot G$ in a multidimensional model
with a Ricci-flat internal space and a multicomponent perfect fluid. A
two-component example, dust + 5-brane, was considered explicitly. It was
shown that $\dot G/G$ is smaller than $10^{-12} yr^{-1}$. Expressions
for $\dot G$ were also considered in a multidimensional model with an
Einstein internal space and a multicomponent perfect fluid \cite{DIKM}. In
the case of two factor spaces with non-zero curvatures without matter, a
mechanism for prediction of small $\dot G$ was suggested. The result was
compared with our exact (1+3+6)-dimensional solutions obtained earlier.

A multidimensional cosmological model describing the dynamics of $n+1$
Ricci-flat factor spaces $M_i$ in the presence of a one-component
anisotropic fluid was considered in \cite{AIKM}. The pressures in all spaces
were supposed to be proportional to the density: $p_{i} = w_i \rho$, $i =
0,...,n$. Solutions with accelerated power-law expansion of our 3-space
$M_0$ and a small enough variation of $G$ were found. These solutions exist
for two branches of the parameter $w_0$. The first branch describes
superstiff matter with $w_0 > 1$, the second one may contain phantom
matter with $w_0 < - 1$, e.g., when $G$ grows with time.

Similar exact solutions, but nonsingular ones and with an exponential
behaviour of the scale factors, were considered in \cite{IKMN} for the same
multidimensional cosmological model describing the dynamics of $n+1$
Ricci-flat factor spaces $M_i$ in the presence of a one-component perfect
fluid. Solutions with accelerated exponential expansion of our 3-space $M_0$
and small variation of $G$ were also found.

Here we continue our investigations of $\dot G$ in multidimensional
cosmological models. The main problem is to find an interval of the
synchronous time $\tau$ where the scale factor of our 3D space exhibits an
accelerated expansion according to the observational data \cite{Riess,Perl}
while the relative variation of the effective 4-dimensional gravitational
constant is small enough as compared with the Hubble parameter, see
\cite{Hel,Dic,DIKM,BZhuk,Zhuk} and references therein.

As we have already mentioned, in the model with two non-zero curvatures
\cite{DIKM} there exists an interval of $\tau$ where accelerated expansion
of ``our'' 3-dimensional space co-exists with a small enough value of
$\dot G$. In this paper we suggest an analogous mechanism for a model
with two form fields and several scalar fields (e.g., phantom ones).

\section{The model}

We here deal with $S$-brane solutions describing two electric branes and a
set of $l$ scalar fields.

The model is governed by the action
\bear{1.1}\lal
   S=\int d^Dx \sqrt{|g|} \biggl\{
    R[g]- h_{\alpha\beta} g^{MN}\p_M\varphi^\alpha \p_N\varphi^\beta
\nnn \cm\cm
    -\sum_{a = 1,2}\frac{1}{N_a!} \exp[2\lambda_a(\varphi)](F^a)^2
        \biggr\}.
\ear
Here  $g=g_{MN}(x)dx^M\otimes dx^N$ is the $D$-dimensional metric of
pseudo-Euclidean signature $(-,+, \dots, +)$, $F^a =  dA^a$ is a form of
rank $N_a$, $(h_{\alpha\beta})$ is a non-degenerate symmetric matrix,
$\varphi=(\varphi^\alpha) \in \R^l$ is a vector of $l$ scalar fields,
$\lambda_a(\varphi)=\lambda_{a \alpha}\varphi^\alpha$ is a linear function,
with $a = 1,2$ and $\alpha, \beta =1, \dots, l$, and $|g| = |\det (g_{MN})|$.

We consider the manifold
\beq{1.2}
    M =    (0, + \infty)  \times M_1 \times M_2 \times
        M_3 \times M_4 \times M_{5}.
\eeq
where $M_i $ are oriented Riemannian Ricci-flat spaces of dimensions $d_i$,
$i = 1,  \dots, 5$, and $d_1 = 1$.

Let two electric branes be defined by the sets $I_1 = \{ 1, 2, 3 \}$ and
$I_2 = \{ 1, 2, 4 \}$. They intersect on $M_1 \times M_2$. The first brane
also covers $M_3$ while the second one covers $M_4$. The first brane
corresponds to the form $F^1$ and the second one to the form $F^2$.

For the world-volume dimensions of branes we get
\beq{1.3}
    d(I_s) = N_s -1 = 1 + d_2 + d_{2 + s},
\eeq
 $s=1,2$, and
\beq{1.4}
     d(I_1  \cap I_2) = 1 + d_2
\eeq
is the brane intersection dimension.

We consider an $S$-brane solution governed by the function
\beq{1.5}
      \hat{H} = 1 + P \rho^2,
\eeq
where $\rho$ is a time variable,
\beq{1.6}
    P = \frac{1}{8}K Q^2,
\eeq
and
 \beq{1.7}
 K = K_{s} =   d(I_s)\bigl( 1+ \frac{d(I_s)}{2-D}\bigr)
      + \lambda_{s \alpha }\lambda_{s \beta} h^{\alpha \beta},
\eeq
$s = 1,2$, is supposed to be non zero. Here
$(h^{\alpha \beta}) = (h_{\alpha \beta})^{-1}$. Thus $K_1 = K_2 = K$.

The intersection rule is as follows:
\beq{1.10}
     d(I_1 \cap I_{2}) = \frac{d(I_1)d(I_{2})}{D -2}  -
     \lambda_{1 \alpha }\lambda_{2 \beta } h^{\alpha\beta}- \frac{1}{2} K.
\eeq
This relation corresponds to  the Lie algebra $A_2$ \cite{IMJ,IMtop}. Recall
that $K_s = (U^s,U^s)$, $s = 1,2$, where the ``electric'' $U^s$ vectors and
the scalar products were defined in \cite{IM11,IM12}  (see also
\cite{IMC,IMJ}). The relations $K_1 = K_2$ and (\ref{1.10}) follow just
from the formula $(A_{s s'}) = (2(U^s,U^{s'})/(U^{s'},U^{s'})$, where $(A_{s
s'})$ is the Cartan matrix for $A_2$ (with $A_{12} = A_{21} = -1$).

We consider the following exact solution:
\bear{1.11}  \lal
    g = \hat{H}^{2 A} \biggl\{ - d\rho \otimes d \rho
        + \hat{H}^{-4 B} ( \rho^2 g^1 + g^2)
\nnn \cm \cm
   + \hat{H}^{-2 B}  g^3  + \hat{H}^{-2 B} g^4 + g^5  \biggr\},
\\ \lal \label{1.12}
    \exp(\varphi^\alpha) = \hat{H}^{ B \lambda_{1}^{\alpha}
        + B \lambda_{2}^{\alpha}},
\yyy  \label{1.13a}
    F^1 = - Q \hat{H}^{-2}
        \rho d\rho  \wedge \tau_1 \wedge \tau_2 \wedge \tau_3,
\yyy  \label{1.13b}
    F^2 = - Q \hat{H}^{-2}
        \rho d\rho  \wedge \tau_1 \wedge \tau_2 \wedge \tau_4,
\ear
where
\bear{1.14A}
      A \eql 2 K^{-1} \sum_{s = 1,2} \frac{d(I_s)}{D-2},
\\ \label{1.14B}
      B \eql 2 K^{-1},
\ear
 $s = 1,2$. Here $\tau_i$ denotes a volume form on $M_i$ ($g_1 = dx \otimes
 dx$, $\tau_1 = dx$). We remind the reader that all Ricci-flat metrics $g^1,
 \ldots, g^5$ have Euclidean signatures.

 This solution is a special case of a more general solution from
 \cite{GIM-flux} corresponding to the Lie algebra $A_2$. It may also be
 obtained as a special 1-block case of $S$-brane solutions from
 \cite{Is-brane}.

\section{Solutions with acceleration}

Let us introduce the synchronous time variable $\tau = \tau(\rho)$
by the following relation:
\beq{2.1}
  \tau = \int_{0}^{\rho} d \bar{\rho} [\hat{H}(\bar{\rho})]^{A}
\eeq
We put $P < 0$, and hence due to (\ref{1.6}) $K < 0$ which implies $A < 0$.
Let us consider two intervals of the parameter $A$:
\bear{2.2i}  \lal
    {\bf (i)} \ \ \ A  < -1,
\\ \lal \label{2.2ii}
    {\bf (ii)}  \ \ -1 < A  < 0.
\ear

In case (i), the function $\tau = \tau (\rho)$ is monotonically increasing
from $0$ to $+ \infty$, for $\rho \in (0, \rho_1)$, where $\rho_1 =
|P|^{-1/2}$, while in case $(ii)$ it is monotonically increasing from 0
to a finite value $\tau_1 = \tau(\rho_1)$.

Let the space $M_5$ be ``our'' 3-dimensional space with the scale factor
\beq{2.3}
      a_5 = \hat{H}^A.
\eeq
For the first branch (i), we get the asymptotic relation
\beq{2.4i}
    a_5 \sim \const\ \tau^{\nu},
\eeq
for $\tau \to +\infty $, where
\beq{2.4n}
       \nu = A/(A+1)
\eeq
and, due to (\ref{2.2i}), $\nu > 1$. For the second branch (ii) we obtain
\beq{2.4ii}
      a_5 \sim \const \  (\tau_1 - \tau)^{\nu},
\eeq
for $\tau \to \tau_1 - 0$, where $\nu  < 0$ due to (\ref{2.2ii}), see
(\ref{2.4n}).

Thus we get an asymptotic accelerated expansion of the 3D factor
space $M_5$ in both cases (i) and  (ii), and $a_5 \to + \infty$.

Moreover, it may be readily verified that the accelerated expansion takes
place for all $\tau > 0$, i.e.,
\beq{2.6}
     \dot{a}_5 > 0, \qquad   \ddot{a}_5 > 0.
\eeq
Here and in what follows  we denote $\dot{f} = df/d \tau$.

Indeed, using the relation $d\tau/d \rho = \hat{H}^A$ (see (\ref{2.1})),
we get
\beq{2.6a}
    \dot{a_5} = \frac{d \rho}{d \tau} \frac{d a_5}{d \rho} =
        \frac{2|A||P|\rho}{\hat{H}},
\eeq
and
\beq{2.6b}
    \ddot{a_5} = \frac{d \rho}{d \tau} \frac{d}{d \rho}
    \frac{da_5}{d \rho} = \frac{2|A||P}{\hat{H}^{2 + A}} (1 + |P| \rho^2),
\eeq
that certainly implies the inequalities in (\ref{2.6}).

Now we consider the variation of the effective $G$. For our model, the
4-dimensional gravitational ``constant'' (in Jordan's frame) is
\beq{2.7}
      G = \const \cdot \prod\nolimits_{i=1}^{4}
            ( a_{i}^{-d_i}) = \hat{H}^{2A} \rho^{-1},
\eeq
where
\bear{2.7a }
        a_1 \eql \hat{H}^{A - 2B} \rho, \cm   a_2 = \hat{H}^{A - 2B},
\nn
    a_3 \eql a_4 = \hat{H}^{A - B}
\ear
are the scale factors of the ``internal'' spaces $M_1, \dots, M_4$,
respectively.

The function $G({\tau})$ has a minimum at the point $\tau_0$ corresponding to
\beq{2.9}
      \rho_{0} = \frac{|P|^{-1}}{1 +4 |A|}.
\eeq
At this point, $\dot G$ is zero. This follows from an explicit relation for
the dimensionless variation of $G$,
\beq{2.11}
       \delta = \dot{G}/(GH) = 2 + \frac{1-|P|\rho^2}{2 A|P| \rho^2},
\eeq
where
\beq{2.12}
       H = \frac{\dot{a}_5}{a_5}
\eeq
is the Hubble parameter of our space.

The function $G({\tau})$ monotonically decreases from $+ \infty$ to $G_0
= G(\tau_0)$ for $\tau \in (0, \tau_0)$ and  monotonically increases from
$G_0$ to $+ \infty$ for $\tau \in (\tau_0, \bar{\tau}_1)$. Here
$\bar{\tau}_1 = +\infty $ for the case (i) and  $\bar{\tau}_1 = \tau_1$ for
the case (ii).

The scale factors $a_2(\tau), a_3(\tau), a_4(\tau)$ monotonically decrease
from 1 to 0 for $\tau \in (0, \bar{\tau}_1)$ since the powers $A-B$ and
$A-2B$ are positive and $P < 0$. The scale factor $a_1(\tau)$ monotonically
increases from zero to $a_1(\tau_2)$ for $\tau \in (0, \tau_2)$ and
monotonically decreases from $a_1(\tau_2)$ to zero for $\tau \in (\tau_2,
\bar{\tau}_1)$, where $\tau_2$ is a point of maximum.

We should consider only solutions with accelerated expansion of our space
and small enough variations of the gravitational constant obeying the
present experimental constraint
\beq{2.10}
       |\delta| < 0.1.
\eeq
Here, as in the model with two curvatures \cite{DIKM}, $\tau$ is restricted
to a certain range containing $\tau_0$. It follows from (\ref{2.11}) that in
the asymptotical regions (\ref{2.4i}) and (\ref{2.4ii}) $\delta \to 2$,
which is unacceptable due to the experimental bounds (\ref{2.10}).  This
restriction is satisfied for a range containing the point $\tau_0$ where
$\delta = 0$.

A calculation of $\dot G$ in the linear approximation near $\tau_0$ gives
the following approximate relation for the dimensionless parameter of
relative variation of $G$:
\beq{2.13}
         \delta  \approx  (8 + 2 |A|^{-1}) H_0 (\tau - \tau_0),
\eeq
where $H_0 = H(\tau_0)$ (compare with an analogous relation in \cite{DIKM}).
This relation gives approximate bounds for values of the time variable
$\tau$ allowed by the restriction on $\dot G$. (For another mechanism of
obtaining an acceleration in a multidimensional model with a ``perfect
fluid'' see \cite{AIKM,IKMN}).

It should be stressed that the solution under consideration with $P < 0$,
$d_1 = 1$ and $d_5 = 3$ takes place when the configuration of branes, the
matrix $(h_{\alpha\beta})$ and the dilatonic coupling vectors $\lambda_a$,
obey the relations (\ref{1.7}) and (\ref{1.10}) with $K < 0$. This is not
possible when $(h_{\alpha\beta})$ is positive-definite, since in this case
$K > 0$. In the next section we will give an example of a setup obeying
(\ref{1.7}) and  (\ref{1.10}) by introducing ``phantom'' fields.

\section{Example}

Let us consider the following example: $N_1 = N_2 = N$, i.e. the ranks of
forms are equal, and   $l = 2$, $(h_{\alpha\beta})=-(\delta_{\alpha\beta})$,
i.e. there are two "phantom" scalar fields. We also put $d_3 = d_4$.  Due to
(\ref{1.3}), $d(I_1) = d(I_2) = N - 1 = 1 + d_2 + d_3$.

Then the relations (\ref{1.7}) and (\ref{1.10}) read
\beq{3.1}
    \vec{\lambda}_1^2 = \vec{\lambda}_2^2 =
        (N - 1) \bigl( 1+ \frac{N - 1}{2-D}\bigr) - K,
\eeq
and
\beq{3.2}
    \vec{\lambda}_1 \vec{\lambda}_2 =
    1 + d_2 - \frac{(N -1)^2}{D - 2} +\frac{1}{2} K,
\eeq
where $K < 0$. Here we have used that $d(I_1) = d(I_2) = N - 1$ and $d(I_1
\cap I_2) = 1 +d_2$.

The relations (\ref{3.1}) and (\ref{3.2}) are compatible since it may be
verified that they imply
\beq{3.3}
    \frac{\vec{\lambda}_1 \vec{\lambda}_2}
        {|\vec{\lambda}_1| |\vec{\lambda}_2|} \in (-1, +1)
\eeq
i.e., the vectors $\vec{\lambda}_1$ and $\vec{\lambda}_2$, belonging to
the Euclidean space $\R^2$, and obeying the relations (\ref{3.1}) and
(\ref{3.2}), do exist. The left-hand side of \eq (\ref{3.3}) gives
$\cos \theta$, where $\theta$ is the angle between these two vectors.

The solution from \sect 2 for the metric and two phantom fields in this
special case reads
\bear{3.11} \lal
    g = \hat{H}^{2A} \biggl\{ - d\rho \otimes d \rho
        + \hat{H}^{-8/K} (\rho^2 g^1 + g^2)
\nnn \inch
    +\hat{H}^{-4/K}(g^3 +g^4) +  g^5  \biggr\},
\\ \lal \label{3.12}
    \exp(\varphi^\alpha)= \hat{H}^{-2 (\lambda_{1 \alpha}
        +\lambda_{2 \alpha})/K},
\ear
where
\beq{3.13}
       A = \frac{4 (N - 1)}{K (D-2)}.
\eeq

The relations for the form fields (\ref{1.13a}) and (\ref{1.13b})
remain the same. Recall that
     $\hat{H} = 1 + \frac{1}{8}K Q^2 \rho^2$, $K < 0$.
The relations (\ref{3.1}) and (\ref{3.2}) for the scalar products of
dilatonic coupling vectors are assumed. Recall that all factor spaces
$(M_i,g^i)$ are Ricci-flat and $d_i = \dim M_i$ with $d_1 = 1$ and $d_5 =3$.
Hence the spaces $(M_1,g^1)$ and $(M_5,g^5)$ are flat.

Here, as in the previous section, the metric (\ref{3.11}) describes an
accelerated expansion of the fifth factor space $(M_5,g^5)$ for $A \neq -1$.
The 4D effective gravitational ``constant'' $G({\tau})$, as a function of
the synchronous time variable, monotonically  decreases from $+ \infty$
to its minimum value $G_0 = G(\tau_0)$ for $\tau \in (0, \tau_0)$ and
monotonically increases from $G_0$ to $+ \infty$ for $\tau \in (\tau_0,
\bar{\tau}_1)$. Recall that $\bar{\tau}_1$ is finite for $ -1 < A < 0$ and
$\bar{\tau}_1 = + \infty$ for $A < -1$. There exists a time interval
$(\tau_{-}, \tau_{+})$ containing $\tau_0$ where the variation of
$G({\tau})$ obeys the experimental bound $(\ref{2.10})$.

\section{Conclusions}

We have considered an $S$-brane solution with two non-composite electric
branes and a set of $l$ scalar fields. The solution contains five factor
spaces, and the fifth one, $M_5$, is interpreted as ``our'' 3D space.

As in the of the model with two non-zero curvatures \cite{DIKM}, we have
found that there exists a time interval where accelerated expansion of
``our'' 3-dimensional space co-exists with a small enough value of
$\dot{G}/G$ obeying the experimental bounds.

Other results on $G$, $G$-dot and $G(r)$ may be found in
\cite{Er88,Mor,M-JP,KHAR,Paris,Taiwan}.

\Acknow
{This work was supported in part by the Russian Foundation for
Basic Research grant Nr. 05-02-17478. VNM is grateful to colleagues from
Observatoire de Paris --- Meudon for hospitality during his visit in
March, 2007.}

\end{document}